\documentclass[conference]{IEEEtran-3}
\usepackage[hidelinks,
			pdftex,
			pdfauthor={Der-Yeuan Yu, Elizabeth Stobert, David Basin, and Srdjan Capkun},
			pdftitle={Exploring Website Location as a Security Indicator},
			]{hyperref}
\usepackage{booktabs} % For formal tables
\usepackage{amssymb,amsmath}
\usepackage{balance}
\usepackage{color}
\usepackage{breakurl}
\usepackage{xspace}
\usepackage{array}
\usepackage{booktabs}
\usepackage{mathtools}
\usepackage{pbox}
\usepackage{enumerate}
\usepackage{xcolor}
\usepackage[keeplastbox]{flushend} % keeplastbox paramter ensures proper indentation of last reference in bib
\usepackage{pifont}
\usepackage{multirow}
\usepackage[flushleft]{threeparttable}
\usepackage{comment}
\usepackage{caption}
\usepackage{subcaption}

\makeatletter
\def\input@path{{../study2/figures/}}
\makeatother

\newcommand{\inlinetitle}[1]{\noindent\textbf{#1}}

\ifCLASSINFOpdf
\else
\fi
\begin{document}
\title{Exploring Website Location as a Security Indicator}

%\author{}


\author{\IEEEauthorblockN{Der-Yeuan Yu,
						  Elizabeth Stobert,
						  David Basin, 
						  and Srdjan Capkun}
		\IEEEauthorblockA{Department of Computer Science, ETH Zurich}
%		\IEEEauthorblockA{\IEEEauthorrefmark{1}{dyu, estobert, basin, capkuns}@inf.ethz.ch}
%		\IEEEauthorblockA{\{dyu, estobert, basin, capkuns\}@inf.ethz.ch}
}


\IEEEoverridecommandlockouts
\makeatletter\def\@IEEEpubidpullup{8\baselineskip}\makeatother
\IEEEpubid{\parbox{\columnwidth}{
	\small
	\vspace{4em}
    NDSS Workshop on Usable Security (USEC) 2018\\
    18-21 February 2018, San Diego, CA, USA\\
%    ISBN 1-1891562-49-5\\
%    http://dx.doi.org/10.14722/ndss.2018.23xxx\\
    www.ndss-symposium.org
}
\hspace{\columnsep}\makebox[\columnwidth]{}}

\maketitle

\begin{abstract}
Authenticating websites is an ongoing problem for users. Recent proposals have suggested strengthening current server authentication methods by incorporating website location as a comprehensible additional trust factor.
In this work, we explore users' acceptance of location information and how it affects decision-making for security and privacy.
We conducted a series of qualitative interviews to learn how location can be integrated into users' decision-making for security, and we designed a security indicator to alert the user to changes in website locations.
We evaluated our tool in a 44-participant user study and found that users were less likely to perform security-sensitive tasks when alerted to location changes. Our results suggest that website location can be used as an effective indicator for users' security assessments.
\end{abstract}


%\keywords{website location; security indicators; user study}

\maketitle

\section{Introduction}

Users' increasing reliance on the Internet for critical services, such as banking, data storage, and communication, highlights the importance of data security and privacy.
However, users can currently do little to ascertain whether their security is ensured and their privacy is respected by online services.

Users' trust in websites is strongly tied to the problem of server authentication, which is currently achieved using public key certificates and browser security warnings.
Unfortunately, research has found that users frequently ignore or bypass related warnings, exposing themselves to online threats~\cite{akhawe2013alice, almuhimedi2014your, sotirakopoulos2011challenges}.
Users also often fail to notice or understand certificate information~\cite{sobey2008exploring, felt2015improving}, which has been partly addressed with improved interface design~\cite{felt2014experimenting, felt2016rethinking, sunshine2009crying, shin2011empirical}.
%As website impersonation attacks increase~\cite{FakeSSLCert, phishingnakedsecurity}, novel mechanisms to strengthen server authentication are being explored.

Recent research on strengthening server authentication has proposed using the server’s geographic location as an additional trust factor~\cite{abdou2016server, yu2016salve}. These proposals integrate web server location into the TLS protocol to increase the security of server authentication, but they require the user to assess and decide whether these locations are trustworthy in situations where the protocol fails (similar to current implementations of certificates). Although there are techniques to display information about server locations, it remains unexplored how such information should be presented and how users would react to it when they make security decisions. 

In this paper, we explore users' decision-making processes regarding their security when they are provided with the location information of websites.
Our goal is to better understand how the proposed website localization techniques will affect users.
%Our goal is to provide insight into users' acceptance of website localization techniques once they are deployed.
Using a user-centered design approach, we gathered requirements, designed a location indicator, and evaluated its usability and effect on security decision making.
 
We conducted semi-structured interviews with 15 participants and applied thematic analysis to identify issues relevant to online trust.
Based on our themes, we developed a model to describe the factors in users' trust and analyzed the role of website locations in their decisions.
We found that users' perceptions of online security and web authenticity are often intermixed with their concerns about privacy.
Our participants often assessed their security on a conceptual level by gauging their risks in terms of financial security and personal privacy.
We also found that participants, while describing various trust concerns about their online security, expressed preferences for particular locations when dealing with sensitive information or transactions.
These findings suggest that the website location is a tangible concept and such knowledge affects users' security decisions.
% knowledge of website locations plays a potential role when users make security decisions.
%trust factor that can be used to address users' mixed concerns for online security and privacy.

% We found that while users did not explicitly refer to website location as a trust factor, they still expressed preferences for website locations when faced with important or sensitive decisions.
% The interviews also allow us to identify users' incentives to maintain online security, how they typically interact with security solutions, and how they incorporate the knowledge of website locations into their trust assessment process.

Based on the qualitative analysis, we derived requirements for a location tool to inform users of website locations.
We designed and implemented \textsf{LocationWatch}, a Chrome extension that makes website locations available to users and alerts them to changes in server locations.
% with the following features: an icon showing the flag of the website's residing country with on-demand information and a warning when the location has changed.
Using \textsf{LocationWatch}, we conducted a user study with 44 participants to analyze how website locations affect their security decisions.
Our statistical analysis showed that participants' decisions were significantly affected by website locations, with fewer users completing sensitive tasks when the website location had changed.
The participants' decisions also varied depending on the sensitivity of data in different application scenarios.
% Our goal is that users can leverage the website location as an additional trust factor to make more informed decisions when accessing important online applications.

The effects of website location knowledge on users' decision-making processes have not been investigated until now.
With recent proposals for strengthening authentication using website locations, it is important to evaluate how this information is perceived by users and how it can be best leveraged in their decision-making processes.
Our results show that users are sensitive to website locations when informed in a non-intrusive way.
This shows the promise of using location information as an additional factor to improve user security and privacy.
% We evaluate the how users' knowledge of website locations affect their security awareness and decision-making processes, which has not been done before.

\section{Background}
\label{sec:background}

Current research on location-based website authentication raises the question of how users might leverage such location information in decision-making. Compared to digital certificates, the tangibility of location and its clear relationship to the real world suggest that location can play a role in users' security and privacy awareness.

% Since location is a more tangible concept than digital certificate validation, information related to it may be more easily consumed by the user to make trust decisions.

\subsection{Location-based Decision-Making}

Psychological research on decision-making has found that people tend to underestimate risk. Since safety and security are abstract concepts, users are unmotivated to pay attention to these risks. Research examining how users make decisions about computer security has found that users reason inconsistently about their gains and losses, and are likely to over-prioritize the cost of losses~\cite{West.2008}. For security, because the gains are abstract and the consequences seem random, users often focus on costs, which are immediate and tangible~\cite{West.2008}. Users may consider gains, such as protecting information, money, and property, but that they are unaware of risks relating to money and property loss online~\cite{Hardee.2006}. Users are also concerned about personal inconvenience in using online services.

Users' security decisions often serve to protect their personal privacy. Research has found that users' privacy preferences are context-dependent and can be easily influenced~\cite{Acquisti.2015}. Users also experience high uncertainty about whether and to what extent they should be concerned about data privacy. Human decision-making can appear inconsistent, but it is governed by a complex calculus of decision-making~\cite{Laufer.1977} that factors in additional information such as social norms and emotional responses. 

Recently, Ruoti et al.~\cite{ruoti2017weighing} explored how people determine the security measures they use to protect their online activities.
They conducted 23 semi-structured interviews with middle-aged residents in Washington state and found that users were often pragmatic about their security decisions due to the appreciation of the convenience brought by the Internet.
They also found that users often have misconceptions of existing TLS security indicators, resulting in insecure behaviors that put their privacy at risk.

Little research to date has analyzed users' perceptions of where their data is stored or to what locations it is transmitted over the Internet. Kang et al.~\cite{kang2015my} conducted a qualitative study investigating users' mental models of the Internet and found that users had only a vague understanding of where data is stored online. They also found that factors such as reputation and appearance were likely to influence users' perceptions of what was happening to their data. 
Ion et al.~\cite{ion2011home} interviewed users about their data privacy awareness and their attitudes about where their online data should be kept. They found that users generally preferred sensitive data to be stored locally than uploaded to cloud storage. They also identified cultural differences that affect users' understanding and preference for their online privacy.
A large-scale study of website credibility~\cite{Fogg.2001} found that websites were more believable when they communicated the ``real world'' aspect of the organization, were professional and easy to use, and included indicators of trustworthiness. It remains unexplored how users might integrate information about the website's location into their evaluation of these environmental cues.

% While existing techniques readily make use of location information, the problem space of how users respond to such website locations remains to be explored. For example, location information may play a role in users' decision-making process when determining whether to trust a website with personal information. Some factors a user may consider include the sensitivity of the online application, the knowledge of the location, the perception of the local government, or the understanding of local legal aspects. Additionally, a user may also be sensitive to changes in a website's location between consecutive visits, due to reasons such as routing policies, data center migration, or potential phishing attacks.

\subsection{Website Location and Authentication}

Websites are currently authenticated using TLS, which requires the server to have a valid X.509 public-key certificate~\cite{clark2013sok}.
The client's browser must validate this certificate upon connection to the server by checking a certificate authority's (CA) signature and other fields.
However, recent incidents have demonstrated the weaknesses of public key infrastructure against a strong adversary~\cite{FakeSSLCert}.
More specifically, attackers have been able to compromise CAs to obtain a fraudulent certificate of an arbitrary website to impersonate it.
Such attacks have been addressed by a wide range of enhancements to TLS authentication, such as Certificate Transparency~\cite{rfc6962}, and pinning~\cite{rfc7469}.
% We refer the reader to a more thorough treatment on TLS authentication by Clark and van Oorschot.
Despite improvements to certificate validation, a strong attacker may still be able to compromise a web server and obtain the private key associated with the public key in its certificate, e.g., by exploiting TLS implementation bugs~\cite{heartbleedcve} or zero-day vulnerabilities.
In these scenarios, server impersonation attacks can be performed remotely by the adversary that controls the network.

Recent work has proposed using website location as an additional authentication factor to strengthen website authentication. 
Verifying a server's location during authentication detects remote server impersonation attacks resulting from the compromise of CAs or websites' private keys.
Yu et al. proposed adding location information to digital certificates to authenticate servers in TLS handshakes~\cite{yu2016salve}. In this approach, a trusted party estimates the location of a website server and issue a signed statement binding the server location to a particular connection with the client. 
The browser can either perform automatic verification (e.g., during the TLS handshake) of the location information or directly display it to the user.
Abdou and van Oorschot proposed similar methods of augmenting TLS by actively estimating website locations using delay-based measurements from multiple locations~\cite{abdou2016server}.
% Both works propose the concept of integrating website location into server authentication and detect location changes.
These approaches leverage the uniqueness and verifiability of private web server locations to supplement existing server authentication.
% As a result, the user can deliberate over the website's authenticity and trustworthiness based on the server's location in addition to its existing digital certificate. 
% is this really something that's handed off to the user? or is there an automatic component

Using location as an authentication factor is increasingly possible due to the availability of pervasive location information, IP geolocation services, and general localization techniques.
A non-technical approach to website localization is the use of public ledgers to record and make available the location of data centers. Online services often host their web servers in data centers, whose locations are publicly known. For example, online resources such as Data Center Knowledge~\cite{datacenterknowledge} provide a public listing of data center deployment and news about web hosting companies. Companies are also increasingly disclosing their server locations to the public~\cite{googleserver, amazonserver}, and using on-site security to protect critical online services from physical intrusion by malicious parties~\cite{googlewhitepaper, visaserver}.
In addition to out-of-band channels, CAs can also verify the locations of online firms and store them in Extended Validation (EV) certificates~\cite{evcerts}, which can be extracted by the browser.
% This allows the browser to learn the legitimate server locations certified by the issuing CA.

Currently, IP geolocation is the most common source of website location data. There also already exist software solutions showing IP geolocation data to users, such as Flagfox~\cite{flagfox} and IP Whois \& Flags~\cite{ipwhois}. However, they do not guarantee that the web servers really are at these locations upon client connection.

In general, with these website location solutions, users are called upon to notice location information and react appropriately. The impact to users' security awareness and decisions has not been explored in depth.

\section{Research Overview}
\label{sec:questions}

Given recent trends in data localization and proposals for location-based authentication, we aim to explore how server location information can be leveraged as a part of users' trust in online services.
%Given the increasing availability of server location information, our goal was to explore how they can be leveraged as part of users' trust in websites and online services.

%Since we aim at evaluating the effect of location verification when presented to users, 

Since we are investigating users' involvement with location information, we inherit the same attacker model proposed in related work on TLS~\cite{clark2013sok} and website location authentication~\cite{abdou2016server, yu2016salve}.
Specifically, we assume that the attacker is able to impersonate the server by compromising its public key certificate, e.g., by obtaining a fraudulent certificate from a compromised CA or learning the server's private key.
We also inherit the assumption that the remote attacker is unable to physically co-locate with the victim's website and resides in a separate location.
The attacker's goals may consist of stealing user data (e.g., passwords, credit card numbers, or personal files) or providing false information (e.g., fake news).
%Concerning the use of website location as an authenticating factor,
We specifically aimed to answer the following research questions about user behavior.
\begin{enumerate}[RQ1]
	\item How do users currently make online security decisions and how could location play a role in these decisions?
	\item Does information about website locations affect users' behavior when they perform online tasks?
\end{enumerate}
% \inlinetitle{Methodology.}
We explored these problems using a user-centered approach~\cite{iso9241:210}. To answer RQ1, we conducted a series of qualitative interviews and applied thematic analysis to understand users' decision-making processes for online security. The themes we identified allowed us to develop a model of users' trust assessments and derive design requirements for a website location tool for a broad range of web users. To answer RQ2, we designed a location tool that displays web server locations, which we implemented as a Chrome browser extension. We conducted a user study to evaluate the usability of our location tool and analyze the impact of location information on users' decisions in real-world application settings. All studies involving human subjects were approved by the ethics committee in our institution.

\section{Study 1: Qualitative Interviews}

We first interviewed users about how they currently determine websites' trustworthiness.
Our goal was to understand how location information could fit into users' decision-making practices and to identify design requirements for a location indicator.
% Because website location has not been widely used as a security indicator in online contexts, there was little on which to base our design requirements.
% Therefore, to better understand users' motivations and understanding of location, we interviewed users about how they currently make decisions in various online scenarios.
% We focused on how location affects their decisions and how it might influence their trust in websites.z

\subsection{Study Design}
We chose a semi-structured interview approach to ensure that we covered topics of interest while giving participants the freedom to discuss their decision-making processes and concerns. Our interview covered three areas: Internet use, security awareness, and location-related preferences. We carefully selected topics that might have associated security or privacy concerns for different Internet usage scenarios: online file storage, emails and calendars, online financial transactions (banking and shopping), and social media. For each topic, we asked about how participants used these services, the kinds of data they stored or obtained through those services, and what kinds of security and privacy concerns they had around these activities. Regarding security awareness, we asked participants about their general security and privacy precautions and where they thought Internet data was stored and served from. Because we were interested in the development of a security indicator, we asked about how they currently determine that websites are legitimate or trustworthy. In the final part of our interview, we explained the concept of website location as a security indicator, and asked participants how they might use it if it were available.\footnote{To make it easier for the participants to understand, we used the term ``website locations'' in our user studies. We use the terms ``website locations'' and ``server locations'' interchangeably throughout the rest of the paper.}
Our interview script can be found in the appendix.

Because using location as a website security indicator is a novel concept, we did not expect participants to explicitly identify it during the interviews. We therefore framed our interview broadly and encouraged discussion on a wide range of topics with relevant security and privacy concerns. By eliciting detailed feedback about users' current decision-making strategies, we sought to understand how location is currently perceived and how it can be used in users' security decisions.
%In our interview design, we made an effort to encourage discussion on a wide range of topics with relevant security and privacy concerns.
Rather than specifically introducing technical concepts of location-based authentication, we introduced topics that naturally led to the subject of location.
If participants did not bring up the subject of location on their own, we attempted to steer the conversation in that direction.

We audio-recorded the interviews to facilitate subsequent note-taking and transcription for analysis.
Participants also completed a brief demographic questionnaire before the interview. Each interview lasted between 30 and 60 minutes.

\subsection{Participants}
% We recruited English-speaking participants from around the area of Zurich, Switzerland. We specifically tried to recruit people of different genders, ages, education levels, occupations, and who came from a diverse set of regions. We deliberately recruited outside the university by posting ads on notice boards around the city, and by advertising our study on expat mailing lists and websites (to target English-speakers). While our sample is likely not representative of the larger population, we do feel confident that a variety of viewpoints were expressed in our interviews, and that the perspectives and experiences expressed in our interviews are in line with the results of similar studies~\cite{kang2015my,fisher2012short}. 

We aimed to represent a diverse array of perspectives and therefore recruited people of different genders, ages, education levels, occupations, and diverse nationalities.
% We deliberately recruited outside our institution by posting advertisements on notice boards around the city, and by advertising our study on mailing lists and websites.
We deliberately advertised outside our institution using public bulletin boards, online forums, and mailing lists.
While our sample is likely not representative of the larger population, a wide variety of viewpoints were expressed in our interviews.
The perspectives and experiences expressed by our participants were in line with the results of similar studies~\cite{kang2015my,fisher2012short}. 

We reached saturation at 15 participants (8 female, 7 male). They ranged in age from 20 to 59, with most (13) aged between 20 and 39 years old. Participants had a variety of educational backgrounds, and their areas of specialty or occupation included social and natural sciences, engineering/informatics, and healthcare. Their occupations included artist, scientist, and student (with 8 students making up the majority). To provide a rough measure of the participants' level of international experience, we asked participants for their nationality, and how many countries they had visited. Participants' nationalities spanned 12 countries, and each participant had visited a median of 10 countries. %(average 11.2)

\subsection{Thematic Analysis}

We reviewed the audio recordings and transcribed each interview. This produced a qualitative dataset that we analyzed using thematic analysis~\cite{Braun.2006}, a flexible qualitative analysis methodology that allowed us to identify themes and relationships in the data. We began our analysis with open coding. We traversed and reviewed the transcriptions line by line and assigned codes to recurring ideas. To ensure consistency, each interview was coded by two researchers, and codes were cross-checked to improve reliability.

An example of our open coding is shown in the following quote, where a participant was asked how she verifies website authenticity: 
% As an example, when asked about how she verified website authenticity, a participant replied:
\begin{quote}
	``I didn't think of [authenticating websites] before. I think every website will give us some legal documents to read before we give information to them. I will scan the documents.''~--~P5
	%   ``Actually I didn't think of [authenticating websites] before. I think every website will give us some legal documents to read before we give information to them. I will scan the documents. I know a lot of people, they don't want to look at it. I also don't want to look at it, but I will just scan. I think it's kind of formal, it's not so fake.''~--~P5
\end{quote}
We assigned the code \emph{lack of awareness} to highlight the participant's lack of concern. Because she mentioned her attention to legal documents, we assigned the code \emph{legal concern}.
We identified 46 open codes in our data. %, and a complete list can be found in Appendix~\ref{sec:opencodes}.
Following the process of open coding, we refined the codes and classified them into themes, described in the subsequent sections. These themes highlight patterns of typical behavior, rather than representing categories of users.

% The findings here allow us to understand users' present knowledge of online security and how location can be used to improve it.

\subsubsection{Trusting by Default}

When asked about their online decision-making, many participants described taking the security of websites for granted without much investigation.
\begin{quote}
	``You just go to the webpage, it looks familiar, and then it never crosses your mind that it may have been forged.''~--~P12
%	``Well, I mean, [website authentication] is not something I think about. You just go to the webpage, it looks familiar, and then it never crosses your mind that it may have been forged.''~--~P12
	% it's
\end{quote}
%This showed a common misconception that the website's authenticity can be intuitively judged by its appearance. 
We also noticed users' default approach to trust in the way they described their automatic use of various online services, such as synchronization of data (e.g., contacts and files) across different devices linked to the same platform.
\begin{quote}
	``I do use sometimes iCloud. I think it just come automatically with my iPhone. Each two weeks, asking  me if I want to store it [...] I just let it.''~--~P1
	%   ``I think I do use sometimes iCloud. I think it just come automatically with my iPhone. Each two weeks, asking  me if I want to to store it... I think mainly I, I just let it. And because I have nothing special.''~--~P1
\end{quote}
%This participant appeared to be unsure of how she used cloud storage and was unconcerned about her profile information being transmitted outside of her device. 
Many participants embraced the convenience of automated functions, such as allowing web email servers to automatically store email addresses of frequent contacts. 

Most participants' initial approach toward online security was to trust that the default configurations are secure. Few participants mentioned looking out for browser security indicators, such as the lock icon or website certificates. When asked about decision-making, participants did not frequently engage in discussions of security and privacy until potential online risks were specifically brought up. Most participants reported using the Internet by simply trusting the way it is.
%I MOVED THIS QUOTE HERE, BUT MAYBE IT JUST OPENS A CAN OF WORMS?
\begin{quote}
	``One keeps hearing about Internet security and all this, but unless something happens, you don't pay a lot of attention to it.''~--~P12
\end{quote}

% There was an evident lack of awareness, which showed there is room for improving browsers to aid users in their trust decisions.
%Some participants also lacked awareness of the potential risks of not properly verifying websites using security indicators such as the lock icon.

\subsubsection{Having Diverse Areas of Concern}
\label{sec:trust_concerns}

% While a few described themselves as generally indifferent, most participants had numerous concerns

Although their default approach was to view the Internet as secure, most participants were able to elaborate areas of specific concern regarding the security and privacy of their data. Among these areas were concerns about personal privacy, financial safety, and freedom of speech.

Personal privacy was a major concern that was brought up repeatedly during the interviews. Participants discussed privacy concerns about sharing information with both online services and other users of those services (and often conflated these two threats).
\begin{quote}
 	``I just kind of like the idea of not being very traceable, not because I'm hiding something specifically but because it's my own business kind of, where I am, what people I'm seeing.''~--~P14
%   ``I just kind of like the idea of not being very traceable, not because I'm hiding something specifically but because it's my own business kind of, where I am, what people I'm seeing. I don't make my life very public...''~--~P14
\end{quote}
Some participants were aware of data collection but were ignoring the implications or did not perceive this as a threat.
%\begin{quote}
%%``As a user, I don't really see the problem. I mean, okay, [online services] are going to have my numbers, and other numbers, but it doesn't really affect me.'' --~P2
%	   ``I don't really see the problem. I mean, okay, [online services] are going to have my numbers, and other numbers, but it doesn't really affect me.'' --~P2
%\end{quote}

% Others mentioned the use of their public profile information for professional networking. 
However, other participants acknowledged the necessity of disclosing personal information. For example, P5 stated that ``sometimes we have to be checked by other people'' (referring to public security). Others regarded the purpose of the Internet as being to share information, and said that curtailing this sharing would render their online presence less meaningful. 
\begin{quote}
 %  ``If [the information is] professional then I put it [online]...  That's not a problem because it actually helps me connect with other people because they know where I worked.''~--~P15
% ``If it's professional then I put it [online]...  If someone knows where I worked, so that's not a problem because it actually helps me connect with other people because they know where I worked.''~--~P15
 ``If someone knows where I worked, that's not a problem because it actually helps me connect with other people.''~--~P15
\end{quote}

A major concern repeatedly mentioned was financial security.
Many participants discussed security concerns around online banking and shopping. 
For example, many participants declined to allow websites to store their credit card information.
%\begin{quote}
%  ``Never, especially after some information in mass media [...] about a risk of misuse''~--~P11
%\end{quote}

Regarding freedom of speech, a few were concerned about unforeseen consequences of disclosing their opinions. %that their public exposure and opinions could lead to future consequences.
\begin{quote}
	``I don't really trust that [my words] might not one day be used against me... a lot of this information is stored and it's just uncomfortable.''~--~P14
	%   ``I don't really trust that [what I say] might not one day be used against me. I don't, I think a lot of this information is stored and it's just uncomfortable I think.''
	%   
	%it's it
\end{quote}

%Some participants take precautionary measures before making online purchases.
%\begin{quote}
%	``I never trust the information even if the website says that it's safe to buy there. I always try to think or to ask friends if they have ever bought something in that website.''~--~P10
%\end{quote}

%When discussing online banking, P8 avoided its use due to a lack of perceived reliability:
%\begin{quote}
%  ``I don't trust the Internet very much in general, and I think everything uses your personal data. Big money transfers through the Internet for me are not very reliable, so I don't want to.''~--~P8 % DO WE HAVE A BETTER QUOTE?
%\end{quote}

% These concerns suggested that many users worry about the risks of using online services, particularly in specialized domains.
%Evidently, there is room for closing the gap between online presence and user comfort.

\subsubsection{Relying on Multiple Trust Factors}
When discussing how they decided to trust websites, participants mentioned a variety of factors. 
% While a few participants were aware of security indicators such as certificate validation, 
Most participants associated website trustworthiness with subjective impressions, such as familiarity of brand presentation, the website interface, and the past experiences of themselves and friends. Even knowledgeable participants admitted to relying on such non-technical cues.
\begin{quote}
	``The first [thing I notice] would be the brand, the logo itself [...] does it look the same?''~--~P2
%		``I think the first [thing I notice] would be the brand, the logo itself [...] does it look the same?''~--~P2 %
	 % For example I think like certain, like, because certain brands is associated with certain fonts, certain color. So if the color scheme is a bit different or the logo is different then that would raise some alarm.''~--~P2
\end{quote}

One major trust factor was the company's reputation. For example, when asked about why they trust particular storage services, some participants relied on the brand name: ``I think having Apple's name behind it, it's quite safe.'' (P3) Participants also listed firms like Google and Amazon as their trusted service providers. Many preferred to avoid unknown shopping websites and rely on payment services with buyer protection policies (e.g., PayPal).

In addition to their own previous experiences, participants also relied on experience from friends or website reviews to judge whether websites were trustworthy. These social cues were used to help discern trustworthiness. 

\begin{quote}
	``[How do you choose where to shop online?] ... usually based on the community. [\dots] I always try to think or to ask friends if they have ever bought something in that website.''~--~P10
%   ``[How do you choose where to shop online?] Um I usually, usually based on the community. I usually read some forums because I never trust the information even if the website says that it's safe to buy there. I always try to think or to ask friends if they have ever bought something in that website.''~--~P10
\end{quote}

\subsubsection{Taking Risks for Practicality}
Most participants described heuristics for decision making based on their trust factors and concerns. These included using pseudonyms, providing fake profile information, avoiding saving credit card information, and only buying from known vendors.

% Users were generally concerned about online exposure and took steps to preserve their privacy and anonymity. When discussing social networking, some participants used pseudonyms and provided fake profile information:
%\begin{quote}
%  ``If I don't know a platform and if I didn't use it before, and I, basically, is it any possibility to use it without my real name, then I usually use without. So I use a nickname, and if I use a nickname I definitely do not give the correct birth date.''~--~P7
%\end{quote}
%People also described behaviors such as avoiding saving credit card information on websites, and only buying from certain vendors. Expert participants referred to mechanisms such as two-factor authentication to improve their security.

% tradeoffs
% 
However, participants often admitted to making exceptions for practical reasons. They justified these decisions by discussing the acceptability or manageability of the potential risks, e.g., a small financial risk when ordering from an untrustworthy merchant. Such decisions often depended on the urgency of the matter at hand.
\begin{quote}
	``If I'm doing stuff on the Internet, I just want it done as fast as possible so I can do something else.''~--~P14
\end{quote}

Compromises were thus often made in the presence of security warnings.
Users put themselves in insecure situations (e.g., by ignoring certificate warnings) to ensure convenience and access to online services. In such situations, participants described a tradeoff between personal security and service accessibility when making their decisions. Though security compromises were made, participants mentioned various secondary measures to reinforce their decisions, such as obtaining tangible proofs of their transactions (``I want to have photocopy or paper as proof''~--P1) or contacting customer service.

\subsubsection{Helplessness and Learning from Consequences}

When discussing their decision-making processes and concerns, participants often expressed frustration over missing information or knowledge that prevented them from behaving securely. Many expressed a kind of learned helplessness relating to their inability to understand security measures.
%\begin{quote}
%	``[How do you know you are visiting the real website?] I don't think too much on these things because I don't know exactly how it works.''~--~P1
%\end{quote}

% ``Company could say one thing and do another thing''~(P2).
Another aspect of this helplessness originated from users' inability to affect corporate policy and their lack of control over where sensitive data is stored.
%\begin{quote}
%  ``Yeah like it's 2016 now. You don't really have power over [where data is stored] anymore.''~--~P9
%%    ``Yeah like it's 2016 now. You don't really have no power over [where data is stored] anymore. This is not for me to... I don't know.''~--~P9
%\end{quote}
One participant told us that ``a company could say one thing and do another thing'' (P2), suggesting their lack of control and distrust in the companies.

% Overall, many participants mentioned that they lack the necessary awareness and technical knowledge of how various security issues are currently addressed in the Internet. For many security-related questions, such as noticing server certificate indicators, many participants showed signs of confusion.

However, though participants expressed a lack of contentment about not being able to control the security of their information, some mentioned that having that control could be a burden to them. 
\begin{quote}
	``If location would be available for me, I would have a feeling that from that time I am the one who has to be responsible for that.''~--~P7
%	``If location would be available for me, I would have a feeling that from that time I am the one who has to be responsible for that also.''~--~P7
\end{quote}

% Efforts to reduce users' knowledge gap by providing more security-related information also implied that users need to be more involved in security assessments. Unfortunately, participants acknowledged the burden of responsibility accompanied by such additional information. More specifically, some participants felt that security obligations were being delegated to them.

% and the helplessness also feeds back into the lousy decision-making and the inconsistencies

We also noticed that some users seemed to have eventually developed a helpless attitude, and described the process of making decisions online as akin to taking a leap of faith: ``I make a wish... I wish nothing happens'' (P6).

\subsection{The Process of Decision-Making }

%The participants varied considerably in how they made trust decisions.
%They mostly trusted websites and online services based on the impression of their reputation.
%When discussing online activities, participants cited financial security and personal privacy as their primary security concerns.
%We also noticed that security and privacy are often jointly discussed by participants, as they often based their security decisions on the potential impact to their personal privacy.
%Their decision-making processes for determining website trustworthiness were composed of typical security practices and pragmatic workarounds when the associated risks were deemed acceptable.

\begin{comment}

\begin{figure}[t]
	\centering
	\includegraphics[width=0.8\columnwidth]{figures/DecisionFunnel}
	\caption{Funnel model of participants' decision making processes. The user considers various supporting and contradicting elements, eventually leading to a decision on whether to trust an online service. The decision results in consequences, e.g., logging in despite a security warning leads to identity theft and helplessness from not understanding a security warning.}
	
	\label{fig:DecisionModel}
\end{figure}

\end{comment}

There was considerable variation in how individual participants made trust decisions in online and real-life scenarios.
Our participants described many elements for decision-making, and they were left to combine these elements into each single decision.
%This tension was often explicit -- when participants described rejecting security for pragmatic reasons, or when trying to find an ``acceptable'' level of risk to assume in a dubious transaction.
We identified a model of how users reach a security-related decision based on various considerations.

The user's decision-making process begins by incorporating the materials used by the user to determine the trustworthiness of a website: their default trust, their varying concerns about security and privacy, the list of factors that give them confidence, their past experiences in similar situations, and the demands of the primary task.
During decision-making, certain elements outweigh others, and the user must obtain a single decision that combines all of their priorities, concerns, and trust. In our interviews, participants seemed unable to give clear descriptions of exactly how they weighted these varying considerations, and it was clear that there was a complex personal calculus that formed each decision~\cite{Laufer.1977}. However, users did often describe the tensions of having to make a single decision from an overload of information (and sometimes, a lack of relevant information). 

Following this decision, its consequences (e.g., improved security, identify theft) may impact not only a single user but also their friends and family as other users look for information to feed into their own decision-making processes. If the user chose not to trust a website, they might have a primary task that remains incomplete, and still be looking for ways to accomplish that task. If they did trust the website, and no security problems result, they may relate that positive experience in user reviews or feedback to other users. In other situations, the exact consequences may be unclear, but the experience of having to make that decision may feed into feelings of a lack of a control or learned helplessness.

\begin{figure*}
	\centering
	\begin{tabular}{p{0.8\columnwidth}p{1.2\columnwidth}}
		\parbox{0.8\columnwidth}{\includegraphics[width=0.85\columnwidth]{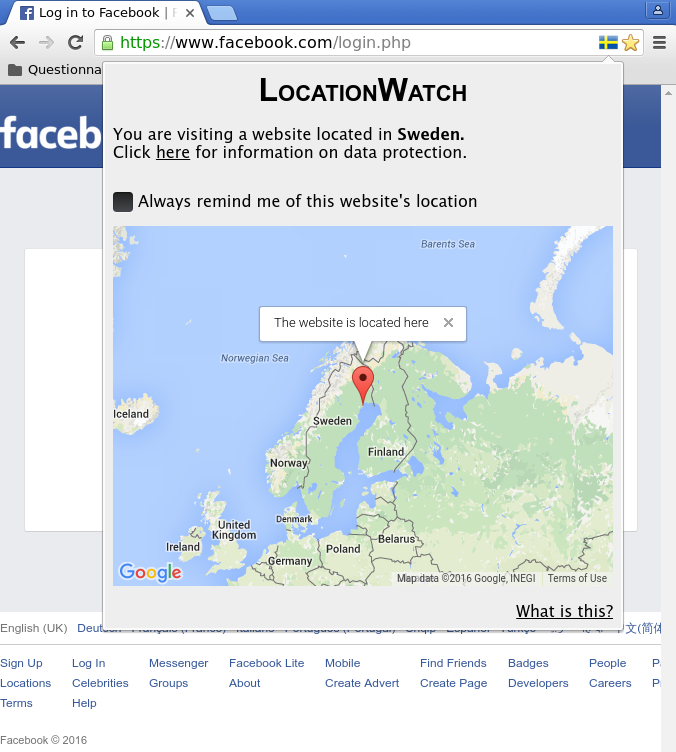}\subcaption{The flag indicator and popup.}\label{fig:snapshot-popup}  }\hfill%
		% The flag of the website residing country in the address bar. Clicking on the flag opened a popup containing further information
		& 
		\begin{tabular}{l}
			\parbox{0.75\columnwidth}{\includegraphics[width=0.75\columnwidth]{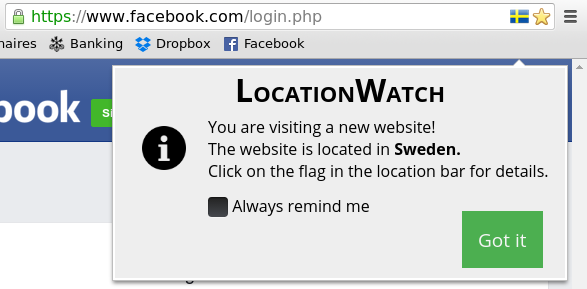}}~%
			\parbox{0.45\columnwidth}{\subcaption{The location tip shown on the initial visit to a website. By default, it is displayed once per website to prevent unecessary obstruction to user experience.}\label{fig:snapshot-tip}}\\\\
			% When the user first visits a website for the first time, the location interface shows a tip information on the website's residing country.
			\parbox{1.10\columnwidth}{\includegraphics[width=1.2\columnwidth]{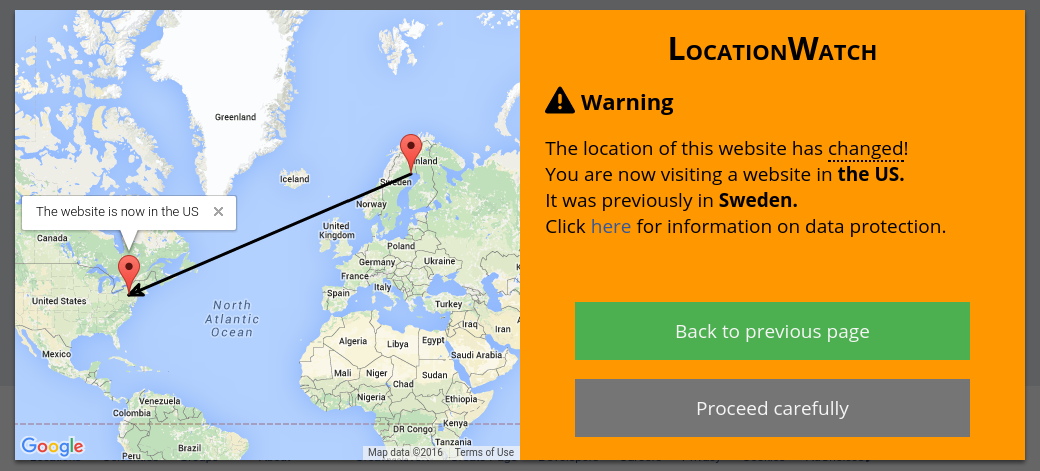}\subcaption{The warning when a website's location has changed}\label{fig:snapshot-warning}}\hfill%
			% When the user revisits a website, the location interface shows an overlay warning over the webpage if the location has changed.
		\end{tabular}
	\end{tabular}
	\caption{Features of \textsf{LocationWatch}, our location indicator.}
\end{figure*}

\subsection{The Role of Website Locations}

In the final segment of our interviews, we briefly explained the concept of location-based website authentication to participants, and asked them for feedback about how they thought it could (or could not) be useful. Users were primed to discuss security in this part of the interview, but since our goal was to design a tool for users, we wanted to understand their desires.

% Given our findings above, we analyzed how location could affect the input to the funnel model and affect user decisions.
Unsurprisingly, participants had not typically related website locations to Internet security. Similar to previous findings~\cite{ion2011home}, they were mostly unaware of the geographic locations where their data was stored. Several speculated that data must be stored in the same countries where the parent companies were based. These responses were sensible and expected since the Internet abstracts away the physical locations of website content and data storage. 
%, while some pointed out online firms' use of multiple data centers for efficiency.
Participants also mostly conflated security (the authenticity of the website) with privacy (where and how users' data is stored or collected). 
%

% We noticed that participants who showed a higher level of default trust were more indifferent to website locations.
% In general, participant also reacted in other ways to website locations, including fear of exposing personal profile to certain governments, something that is ``very interesting to know''~(P5).
% The participants primarily discussed different locations on the country level.

When asked about the presentation of location information, most participants discussed the idea of location on the country level (as opposed to the city or continent level). Participants often brought up the legal implications of having data stored in different countries (mainly in the context of financial information). They also occasionally referred to public disclosures of nation level surveillance programs (e.g., mass surveillance in the USA) and other data-gathering concerns when discussing where they avoided sharing or storing personal data.
% full quote - edited out the reference to Switzerland
\begin{quote}
	``It is important to be sure they are stored in countries with high security levels... legal regulations [on] who is allowed to have access and under what conditions someone could have access to such data. And in Europe, I would say such [legal institutions] are on a high standard.''~--~P11
	%   ``I would say it is important to be sure they are stored in countries with high security levels... I would say in the first place legal regulations: who is allowed to have access and under what conditions someone could have access to such data. And in Europe, I would say such legal institutions legal issues are on a high standard. I would prefer to be sure that such data are stored in such places, such countries.''~--~P11
\end{quote}

% Although many participants were unable to envision exactly how they would leverage verified location information to establish trust, they 
% A few participants were able to articulate ways in which location already influences their trust decisions.
% Some participants expressed clear preferences for certain countries or regions due to their reputable privacy regulations and law enforcement (mainly in the context of financial transactions). 

% Some participants also remarked that the knowledge of website locations do not technically imply the existence of data protection or security mechanisms. One participant argued that since cloud services providers often choose data center locations for practical reasons, on-site security plays a more important role.

%\begin{quote}
%  ``I wouldn't really care that much about the location of the server. Because at the end it's just like a virtual machine located somewhere. I will check more then the, then the security of the site itself, not where it's located... most of the times you have like a lot of nodes in the cloud solutions so it's, they will go perhaps to cheaper countries, so for me it makes sense. They are not going to store here in Switzerland.''~--~P4
% \end{quote}

Participants suggested that location can be incorporated into existing security mechanisms for critical applications.
\begin{quote}
	``I think in the end it will be used everywhere because it would be like an adapted protocol. I think for me it would be useful [in] banking.''~--~P4
	%   ``But I think in the end it will be used everywhere because it would be like a, like an adapted protocol. I think for me it would be useful into banking, into this kind of banking scenarios where there is sensitive data. I would try to keep track of these kinds of location.''~--~P4
\end{quote}

Participants often related website location to the implication of disclosing their data to the foreign governments.
Aside from security concerns due to server impersonation, participants cared about how foreign governments could harm their privacy when websites are hosted abroad.

% Users often drew from past experiences when indicating their preferences for website locations. Most participants indicated explicit preferences for personal data not to be stored in countries with weak privacy laws, or where there had been revelations of mass surveillance. Many users were often admittedly also under the influence of cultural biases, which could have been due to the global reputation or personal stereotypes of countries.

Regarding trust factors, we noticed that participants were more receptive to discussions of website locations as opposed to traditional security solutions, such as public key certificates or authentication. Users had opinions about locations, and were willing to discuss how they might relate website locations to other information about countries or services. One participants discussed the utility of location information in relation to logistical concerns such as shipping and postage when shopping online. Another participant wanted to be able to consider environmental implications of web server locations, and discussed her concerns about the damage inflicted by large heat-generating data centers. Another participant mentioned a similar concern, but related her desire to know that her data was being stored in a location where workers were being treated and paid fairly.

% This showed that there is a gap between the desired goals of current security indicators and their effects on users' decision making. It also suggested the potential of using website location to improve user security awareness and decision making. To evaluate users' perception of and response to such information, we required an assistive tool to display website locations.

\begin{table*}
	\small
	\centering
	\begin{threeparttable}
		\centering
		\begin{tabular}{cllcc}
			
			\toprule
			&&& \multicolumn{2}{c}{\bf Available indicators} \\
			\cmidrule(r){4-5}
			{\bf Stage} & {\bf Tasks} & {\bf Website location} & {\bf Ctrl features} & {\bf Expt features} \\
			\midrule
			1 & Dropbox: upload passport scan & United States & Flag & Flag + Tip \\
			Initial visit & Facebook: update status & Sweden & Flag & Flag + Tip \\
			& Banking: check 1$^\text{st}$ account balance & Switzerland & Flag & Flag + Tip \\
			\midrule
			2 & Dropbox: upload password list & United States & Flag & Flag (+ Tip*)\\
			Re-visit & Facebook: update status & Sweden & Flag & Flag (+ Tip*) \\
			without change& Banking: check 2$^\text{nd}$ account balance & Switzerland & Flag & Flag (+ Tip*) \\
			\midrule
			3& Dropbox: upload credit card & China & Flag & Flag + Warning \\
			Re-visit & Facebook: upload party photo & United States & Flag & Flag + Warning \\
			with change& Banking: check 3$^\text{rd}$ account balance & Japan & Flag & Flag + Warning \\
			\bottomrule
		\end{tabular}
		
		\begin{tablenotes}
			\item *The tip is only shown if the participant checked the ``always remind me'' option in Stage 1.
		\end{tablenotes}
		
	\end{threeparttable}
	\caption{Study 2 tasks and location configuration for the control and experiment groups.}
	\label{tab:study2stages}
\end{table*}

\section{Design of LocationWatch}

Based on our qualitative analysis, we developed a set of design requirements for our location tool, \textsf{LocationWatch}.
\textsf{LocationWatch} is intended to act as a visual indicator to inform users about the result of server location verification, envisioned to use recently-proposed methods~\cite{abdou2016server, yu2016salve}.
We implemented \textsf{LocationWatch} as a Chrome extension featuring a flag indicator, a location tip, and a warning message.
In real-world deployment, \textsf{LocationWatch} would be incorporated into existing security indicators; in this paper, we implemented a prototype as a browser extension.
% to prevent priming the user study participants of the experiment goal.
Before website location verification methods are widely deployed, \textsf{LocationWatch} can use IP geolocation databases as a reference.

\textsf{LocationWatch}'s main features include a flag indicator, a location tip, and a warning message.
The flag indicator (Figure~\ref{fig:snapshot-popup}) is an icon near the address bar showing the flag of the server's residing country.
It also shows more information in a popup window when clicked by the user.
The location tip (Figure~\ref{fig:snapshot-tip}) is a small window on the upper-right corner of the web content that appears on the first visit to a website.
The warning message (Figure~\ref{fig:snapshot-warning}) appears when a website's location has changed since the user's previous visit and allows the user to decide whether to continue visiting it.
In the event of a server impersonation attack (i.e., using a fraudulent certificate or a compromised server's private key), this tool would display the location of the attacker's server.

\subsection{Design Rationale}

We aimed to implement \textsf{LocationWatch} as an unobtrusive and effective tool to assist users in assessing the inputs to their decision.
%Based on our funnel model (Figure~\ref{fig:DecisionModel}), we identified its implications on our design choices.
%We developed \textsf{LocationWatch} as a separate browser indicator so that users can independently evaluate website locations without being explicitly aware of its purpose for security.
We discuss its potential integration with existing security indicators in Section~\ref{sec:discussion}.

\inlinetitle{Default Trust.}
%Participants often said they trusted online services by default and neglected to pay attention to security indicators.
Since users often trust websites by default and without understanding security indicators, security information should be made intuitive for them.
Some may even prefer not to be bothered with location details since website security is not their primary task.
We therefore designed \textsf{LocationWatch} to be non-intrusive by showing only the flag icon by default.

\inlinetitle{Diverse Concerns.}
Participants were often concerned about how their data was used or misused by governing nations in which the web servers reside.
Since legal protection laws differ across countries, the location of where data is stored or sent may prompt different user concerns and influence subsequent decision-making.
We therefore designed a popup (Figure~\ref{fig:snapshot-popup}) that appears when the user clicks on the flag icon.
This popup shows the server's governing country and information on that country's data protection laws for the users' reference.
%The location tip acts as a visible notification for the user on the first visit.
%Users can also opt-in to see this tip for every visit.

% \begin{figure*}[t]
%   \centering
%   
%   \begin{tabular}{cc}
%     \multicolumn{1}{m{0.7\columnwidth}}{
%       \begin{subfigure}[b]{0.7\columnwidth}
% 	\centering
% 	\includegraphics[width=\columnwidth]{figures/snapshot-popup-crop2.png}
% 	\caption{The flag of the website residing country in the address bar. Clicking on the flag opened a popup containing further information.}
% 	\label{fig:snapshot-popup}      
%       \end{subfigure}
%     }
%     & 
%     \begin{tabular}{c}
%       \begin{subfigure}[b]{0.7\columnwidth}
% 	\centering
% 	\includegraphics[width=\columnwidth]{figures/snapshot-tip-crop2.png}
% 	\caption{When the user first visits a website for the first time, the location interface shows a tip information on the website's residing country.}
% 	\label{fig:snapshot-tip}
%       \end{subfigure}\\
%       \begin{subfigure}[b]{1.2\columnwidth}
% 	\centering
% 	\includegraphics[width=\columnwidth]{figures/snapshot-warning-crop.png}
% 	\caption{When the user revisits a website, the location interface shows an overlay warning over the webpage if the location has changed.}
% 	\label{fig:snapshot-warning}
%       \end{subfigure}
%     \end{tabular}
%   \end{tabular}
%   \caption{Features of \textsf{LocationWatch}, our location indicator}
% \end{figure*}

\inlinetitle{Trust Factors.}
%While recent work has shown the potential of location as a trust factor, 
Most participants did not initially think of location as a trust factor.
To strengthen users' attention to location, the location tip appears on the user's initial visit to a website (Figure~\ref{fig:snapshot-tip}).
While slightly obtrusive, this tip provides an attentive user a first impression of where this website is originally located and it is designed to only appear once by default.
We also use the popup window (Figure~\ref{fig:snapshot-popup}) to show more detailed information for interested users.

\inlinetitle{Past Experience.}
Previous experience plays an important role since many participants considered the visual familiarity of websites as a primary factor for trust.
Therefore, we chose to show a visual cue to inform the user when the website location has changed.
This is realized using a warning message (Figure~\ref{fig:snapshot-warning}) showing the current and previous website locations.

\inlinetitle{Practicality.}
Participants admitted to bypassing warnings for practical reasons such as convenience or an acceptable level of risk.
Our location indicator does not prohibit such choices, similar to certificate warnings.
In the warning message, we provided two buttons: ``leave the website'' and ``proceed carefully'' (Figure~\ref{fig:snapshot-warning}).
%The phrasing of these buttons was designed to hint that users should consider he security of their decisions.

%\subsubsection{Helplessness} Participants were often frustrated with technical security measures due to their lack of knowledge and control. The location indicator should have obvious signs, simple and intuitive explanations, and clear actions for the user to take. As a result, \textsf{LocationWatch} contains only simple textual information with a link to detailed legal reference. Users can control when the warning message is shown, allowing them to decide whether and to what extent to consider location information.

%\subsubsection{Location as a Factor} Participants often expressed preferences for certain countries to store their data. This also suggested that country information is a suitable level of detail for the indicator to show to most users. Since some users may want more information about data center locations, detailed locations should also be available. We support this by showing the detailed location using a map in the popup window (Figure~\ref{fig:snapshot-popup}).

\section{Study 2: User Evaluation}

We conducted a user study to evaluate the impact of website location on users' decision-making.
First, we aimed to evaluate how users' security behavior changes when website locations are provided.
Second, we aimed to evaluate the usability of \textsf{LocationWatch} to see if it satisfied our design concepts and requirements. 
%Positive user feedback would suggest the potential of location as useful added information for users. 
% More specifically, we hypothesize that, similar to the interviews, users' decisions would be affected by users' subjective impressions of different countries.
Since the interviewed participants mostly relied on experiences and impressions, we hypothesized that website location changes across subsequent visits would affect users' decisions.

\subsection{Study Design}

To evaluate \textsf{LocationWatch} and users' response to website locations, we designed an experiment where participants used three web services (file storage, social networking, and online banking) and performed routine but potentially sensitive tasks. We chose these services to prompt typical concerns from the qualitative analysis: personal privacy, identity safety, and financial security. We aimed to measure how online behavior varied when participants were given website location information using \textsf{LocationWatch}.

Our study had a mixed design, where group was a between-subjects factor and stage was a within-subjects factor. There were two groups: control and experiment. In the control condition, the location interface was configured to show only the flag icon and the popup window (making it similar to existing tools~\cite{flagfox, ipwhois}). In the experiment condition, participants used the fully-featured version of \textsf{LocationWatch}, including the location tip and the location change warning. The study had three stages and in each stage the participant was asked to perform three tasks, as shown in Table~\ref{tab:study2stages}. We used a Latin square design to shuffle the task order across different participants in each stage to avoid order effects.

% The participant was provided with a laptop to use online services through a browser. We created user accounts for the participants for each website and also provided some personal files stored on the desktop. For file storage, we created a Dropbox account; for social networking, we created a Facebook account; for banking, we used a full-featured demo provided by a local bank. The files we created included: a passport scan, a credit card scan, a password list, and a party photo. 

% \subsection{Procedure}
Each participant was given a brief introduction to the study's purpose as a usability evaluation of a software tool. All participants received the same tutorial on \textsf{LocationWatch}, introducing the concept of geographic locations of websites, the flag icon, and the popup features. To avoid priming users to expect location changes, we did not introduce the location tip and warning (only visible to the experiment group). Participants were then given login information for the accounts and files created for the study, and instructed to treat them as if they were their own.
%\textsf{LocationWatch} was pre-installed in the participants' browser to prevent its installation process from becoming a confounding factor to user response.

%These tasks represent daily activities such as uploading personal files to online storage, posting updates to social media, checking account balances. For each task, the participant had to log in to complete it and log out afterwards. The tasks were designed so that they are short, but significant in terms of personal security and privacy.

% In Stage 1, the participant was asked to upload the passport scan to Dropbox, post a status update disclosing his current location on Facebook, and check the balance in one of his accounts. In Stage 2, the participant was asked to upload the password file to Dropbox, post a status update disclosing his recent activity on Facebook, and check the balance in another account. In Stage 3, the participant was asked to upload the credit card scan to Dropbox, post the party photo on Facebook, and check the balance in yet another account.

% Stage 1 represents the user's first visit to the websites when their locations were shown for the first time: Dropbox in the US, Facebook in Sweden, and the online bank in Switzerland. These are plausible locations in our region since Dropbox stores files in the US~\cite{dropboxstorage}, Facebook has data centers nearby~\cite{facebooksweden}, and the local bank often store data domestically~\cite{swissdatacenter}. Stage 2 represents the second visit where the locations remain the same. In Stage 3, the website locations were changed. 

\subsection{Selecting Test Locations}

To evaluate user reactions to various locations, we programmed fake locations to be displayed by \textsf{LocationWatch}.
We configured our tool to show three types of locations: countries associated with good privacy impressions (Sweden, Switzerland, Japan), a developed country with prominently reported data privacy breaches (the USA), and a developing country with known Internet censorship (China).
For the last stage, we programmed \textsf{LocationWatch} to simulate location changes: Dropbox from the USA to China, Facebook from Sweden to the USA, and the online bank from Switzerland to Japan. For the control group, this led to a change of the country flag and popup contents. For the experiment group, the location change warnings were additionally shown. %somehow describe the idea that the changes are unusual, and then going to a slightly dubious country and then to a very dubious country.

Any choice of countries would naturally subject our study to various user-side cultural biases, and we therefore fixed the country assignments across different participants rather than randomizing them to minimize experiment variation.
Since we were focused on observing whether location plays a role at all, we leave the design of a more large-scale and ecologically valid study as future work.

Each session lasted between 30 and 60 minutes. In addition to the instrumented measurements about their activities, the participants completed three questionnaires: a demographic questionnaire, a pre-test questionnaire about their online decision-making habits, and a post-test questionnaire about their impressions of \textsf{LocationWatch}.
%In addition, we asked the users how they felt when they saw the locations and when locations changed. We also interviewed them about their overall impressions and suggestions.

% \begin{table}[t]
%   \centering
%   \begin{tabular}{lcrrr}
%   \toprule
%   Condition & Stage & Mean & Median & SD \\
%   \midrule
%   \multirow{3}{*}{Control} & 1 & 2.82 & 3 & 0.50 \\
%                         & 2 & 2.55 & 3 & 0.60 \\
%                         & 3 & 2.32 & 3 & 0.84 \\
%   \midrule
%   \multirow{3}{*}{Experiment} & 1 & 2.73 & 3 & 0.63 \\
%                         & 2 & 2.32 & 2 & 0.65 \\
%                         & 3 & 1.45 & 2 & 0.96 \\
%   \bottomrule
%   \end{tabular}
%   \caption{Descriptive statistics of task completion across different stages.}
%   \label{tab:complete_descriptive_stage}
% \end{table}

\subsection{Participants}
%Given the task design, we required users with an existing habit of actively using web services. We recruited participants by posting advertisements on the university campus, online forums, and using a study recruitment platform. 

We recruited users who were aged 18 years or above, spoke English, and had Internet experiences, including online banking, file storage, and email. 44 participants completed the study (23 female and 21 male), most of whom were students (32). They ranged in age from 20 to 59, with most (34) being between 20 and 29 years old. Participants' nationalities spanned 17 countries and they had visited a median of 15 countries.
% including Albania, Argentina, Austria, Balarus, Brasil, China, Colombia, Germany, Greece, Hong Kong, Italy, Macedonia, New Zealand, Singapore, Switzerland, the Netherlands, and the US. 
They come from various backgrounds, including social sciences, humanities, natural sciences, and engineering.
Each study lasted between 25 and 40 minutes.

\begin{figure}[t]
	\centering
	%   \begin{subfigure}[b]{0.5\columnwidth}
	%     \includegraphics[width=\columnwidth]{complete_boxplot_stage_ctrl-crop}
	%     \caption{Control Group}
	%   \end{subfigure}%
	%   \begin{subfigure}[b]{0.5\columnwidth}
	%     \includegraphics[width=\columnwidth]{complete_boxplot_stage_expt-crop}
	%     \caption{Experiment Group}
	%   \end{subfigure}%
	\includegraphics[width=\columnwidth]{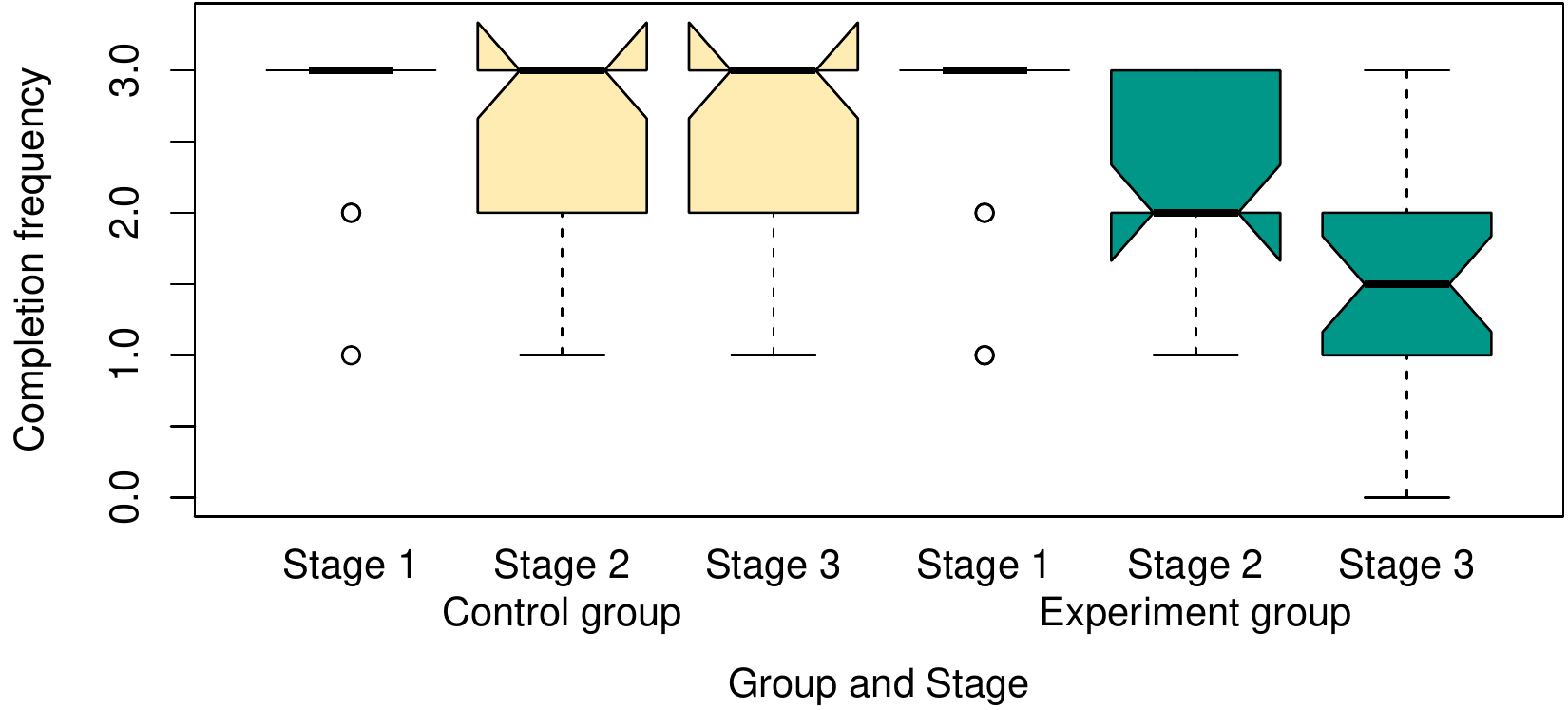}
	\caption{Box plots of task completion scores across different stages for all websites.\protect\footnotemark}
	\label{fig:complete_boxplot_stage}
\end{figure}
\footnotetext{The notches in the box plots represent the 95\% confidence intervals around the median. When the intervals fall outside the 1$^\text{st}$ or 3$^\text{rd}$ quartiles, the notches extend beyond the box.}

\subsection{Results}

We evaluated participant behavior based on the number of completed tasks and the decision-making time.

%We evaluated performance in our study with two main measures: the number of tasks that were completed, and the time that users took in the decision-making process. 

\subsubsection{Task Completion}
We recorded how often users completed the tasks in each stage and each condition of the experiment, and used task completion as a measure of how location affected participants' behavior.
We defined task completion as having logged into the web service \emph{and} completed the given task. We encoded completed tasks as 1, and uncompleted tasks as 0. For each stage we summed the scores from the three websites to produce an aggregate score between 0 and 3. 

Figure~\ref{fig:complete_boxplot_stage} shows the distributions of completion scores. While most control group participants completed all tasks in all stages, fewer experiment group participants completed the task when the location changed. Descriptive statistics are detailed in Table~\ref{tab:complete_descriptive_stage} in the appendix.

We first looked for differences in task completion between the control and experiment groups. Since task completion was based on counts, we performed a between-subjects Chi-squared test on the sum of completion scores across all stages and found a significant difference between the two conditions ($\chi^2(1)=9.44,p=0.002$)%
. Post-hoc pairwise Chi-squared tests using a Bonferroni correction showed that this difference occurred in Stage 3 ($\chi^2(1)=10.52,p=0.011$)%
, where the warning made participants in the experiment group less likely to complete the task.

In the absence of an omnibus test for categorical data, we conducted Chi-squared tests to look for differences between the stages in each condition. We found a significant effect of stages in the experiment group ($\chi^2(2)=30.86,p<0.001$)%
, but no effect in the control group ($\chi^2(2)=7.35,p=0.228$)%
. Table~\ref{tab:complete_chisq_stage} in the appendix shows the results of post-hoc pairwise Chi-squared tests. We found significant differences in task completion between Stage 1 and Stage 3 ($\chi^2(1)=26.15,p<0.001$)%
, and between Stage 2 and Stage 3 ($\chi^2(1)=10.52,p=0.011$)%
 for the experiment group, showing that the warning for location changes significantly affected whether participants completed critical tasks.

\begin{figure}[t]
	\centering
	\includegraphics[width=\columnwidth]{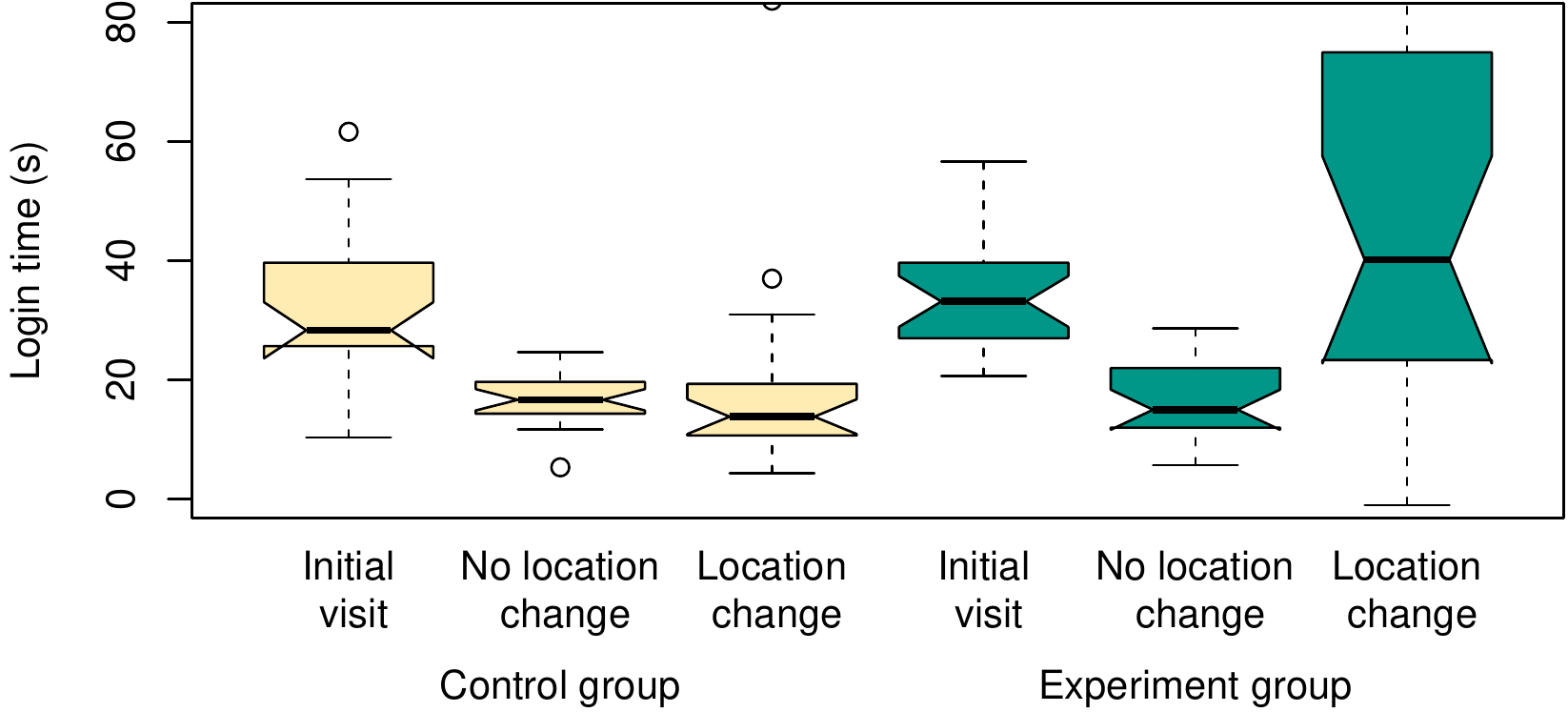}
	
	\caption{Box plots of time spent deciding to log into websites, averaged over all three websites in each stage.}
	\label{fig:login_times}
\end{figure}

\subsubsection{Decision-Making Times}
As an indication of how much attention participants paid to making decisions about location, we recorded the time taken in the login process. We measured the time between when the webpage loaded and when the user clicked the login button (in seconds). This measurement included the time that the user spent deliberating about whether to login. 
%DO WE ONLY MEASURE FOR TIMES WHEN THEY ACTUALLY LOGGED IN???? 
We aggregated the times for each participant in each stage by taking the mean of the times for the three websites. Figure~\ref{fig:login_times} shows the distribution of times across the three stages. 

Table~\ref{tab:login_times_descriptive} in the appendix shows descriptive statistics of the times that participants took to log in by group and stage. The times in Stages 1 and 2 were similar across the two groups, the times decreased in Stage 2 (from $\sim$30 seconds to $\sim$17 seconds). In Stage 3, the experiment group spent more time considering their login decision (54 seconds).

We used a mixed two-way ANOVA to analyze the differences in login times between the two conditions and between the stages. There were significant effects of both condition ($F(1,41)=12.73, p<0.001$) and stage ($F(2,82)=9.92, p<0.001$) and a significant interaction between condition and stage ($F(2,82)=10.65, p<0.001$).

We then used post-hoc pairwise $t$-tests to examine the differences between the two groups.
There were significant differences only in Stage 3 ($t(21)=-3.08,p=0.051$)%
, implying that the warning made the experiment group spend more time than the control group.

We further conducted post-hoc pairwise $t$-tests with a Bonferroni correction to look for differences within each group (Table~\ref{tab:login_times_ttest} in the appendix).
There were significant differences between Stages 1 and 2 for both conditions ($t(21)=6.01, p<0.001$ for control, $t(21)=7.24, p<0.001$ for experiment), possibly because the participants got used to the login process.
The experiment group had significantly different login times between Stages 2 and 3 ($t(21)=-3.43, p=0.023$), implying that the warning affected their time spent deciding to log in.

% % ANOVA TABLE FOR LOGIN TIMES!!!
% % Table contents copied from \input{anova_cond_session.tex}
% \begin{table}[h]
%   \centering
%   \caption{ANOVA table for between-subject and within-subject tests on the login times for each stage across different groups (control and experiment).}
%   \begin{tabular}{lrrr}
%     \toprule
%     & df & F value & Pr($>$F) \\ 
%     \midrule
%     Group      & 1 & 12.73 & 0.0009 \\ 
%     Residuals & 41 &  &  \\ 
%     \midrule
%     Stages      & 2 & 9.92 & 0.0001 \\ 
%     Group:stage & 2 & 10.65 & 0.0001 \\ 
%     Residuals    & 82 &  &  \\ 
%     \midrule
%     Residuals1 & 258 &  &  \\ 
%     \bottomrule
%   \end{tabular}
%   \label{tab:login_times_anova}
% \end{table}

\subsubsection{Task Completion on Different Websites}

%To examine whether \textsf{LocationWatch} had an effect on participants' willingness to complete security-sensitive tasks,
We aggregated completion scores within the same stage in our initial task completion analysis. 
Here, we performed an exploratory analysis to investigate how users reacted to different location changes on different websites.
The scale of our study prevented us from exhaustively testing different websites and locations.
However, in our study design we attempted to pick security-sensitive websites, and to choose location changes that might represent different attacks. 
We included location changes to countries that were neutral but implausible (Switzerland to Japan, banking), locations with well-publicized privacy issues (USA to China, Dropbox), and changes between plausible locations (Sweden to USA, Facebook). 
Our exploratory analysis evaluates whether there was an effect of website (and the corresponding country change) on task completion.

% The tasks reflected different aspects of users' online security concerns as explored in the qualitative analysis, such as financial security and personal privacy.
% To identify different user behavior across different security concerns, we analyzed the differences in task completion across different websites.

\begin{figure}[t]
	\centering
	\includegraphics[width=\columnwidth]{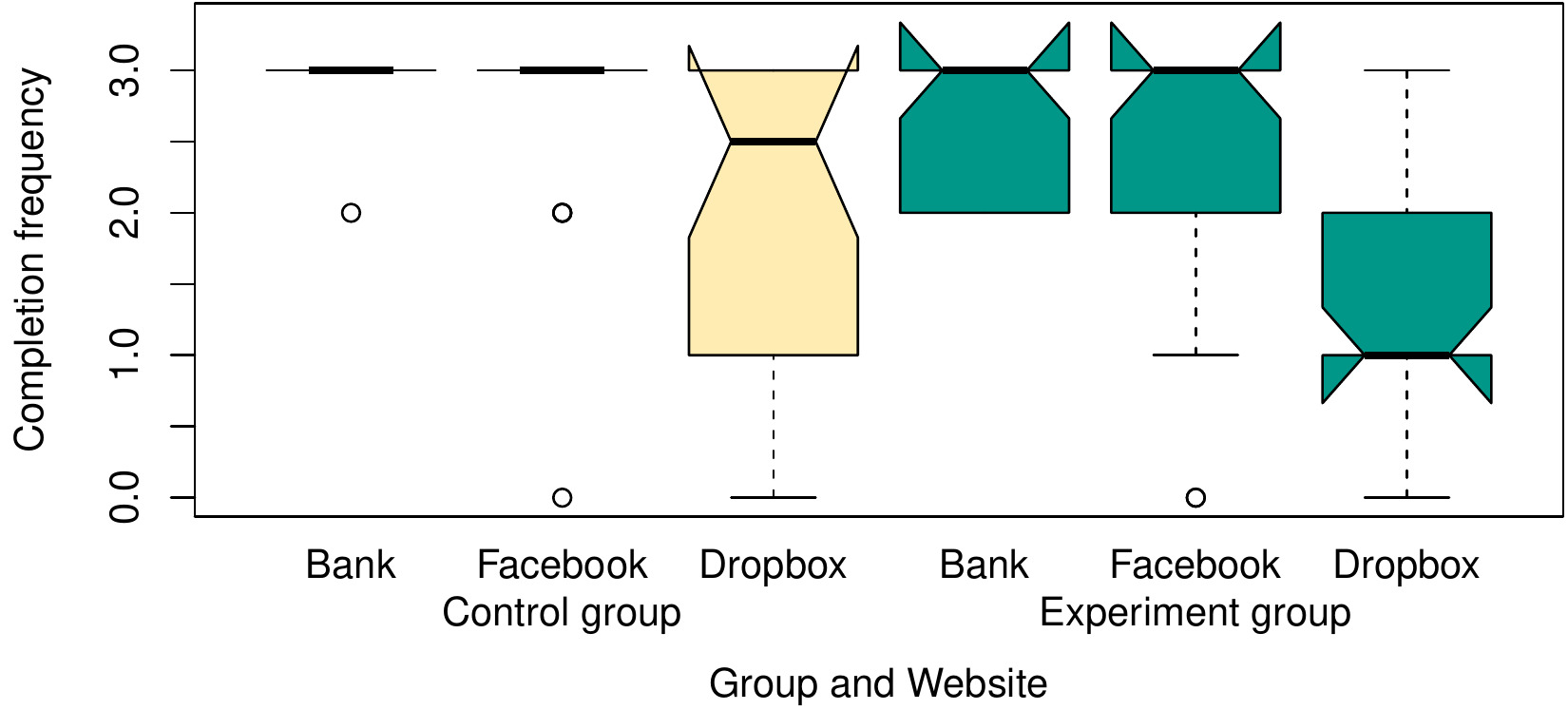}
	\caption{Box plots of the task completion scores across different websites for all stages.}
	\label{fig:complete_boxplot_website}
\end{figure}

% We measured how often users completed the tasks on the same website throughout the entire experiment.
%Using the same definition for task completion as previously, 
Similar to analyzing task completion across stages, we defined a participant's task completion score for each website, ranging between 0 (no tasks completed) and 3 (tasks completed in all websites). The distributions of website task completion scores for the control and the experiment groups are shown in Figure~\ref{fig:complete_boxplot_website}. In both conditions, participants completed fewer tasks on Dropbox (Table~\ref{tab:complete_descriptive_website} in the appendix).

A Chi-squared test using a Bonferroni correction showed a significant effect of website in both the control condition ($\chi^2(2)=29.17,p<0.001$)%
 and the experiment condition ($\chi^2(2)=32.07,p<0.001$)%
. Post-hoc pairwise tests revealed that in both conditions, significantly fewer participants completed the tasks on Dropbox than on Facebook or banking, as shown in Table~\ref{tab:complete_chisq_website} in the appendix.

\subsubsection{Usability}

We used the System Usability Scale~\cite{brooke1996sus} to evaluate the usability of our location interface.
% (detailed SUS results are presented in Appendix~\ref{sec:usability_sus}).
Both variants of our interface were ranked as ``excellent'' (scores greater than 80)~\cite{ruoti2016standard}. The average scores were 81.61 for the control group and 82.2 for the experiment group. A Mann-Whitney test showed no significant difference in usability between the two versions ($U=254.5, p=0.78$).

\subsection{Summary}

Participants' decisions were indeed affected by their knowledge and perception of the websites' locations. There were statistically significant behavior changes in task completion when website locations changed in the experiment condition. This suggested that participants noticed the changes and some avoided sensitive tasks. Participants who completed the tasks despite location changes mainly cited the website's reputation as the main reason.
%Some took extra steps to validate the websites' authenticity, such as connecting to the designated website by searching for it using search engines.

We also noticed that when warned of website location changes, two users in the experiment group still signed in to inspect the website (Facebook and Dropbox) before refusing to perform the given task.
This suggests that for some users, the decision point for personal security or privacy lies past the sign-in process.
This supports previous work on browser warnings~\cite{shin2011empirical} that prevent users from leaking personal credentials before ascertaining their trustworthiness. It also suggests that security indicators could be more useful if they were presented right before critical tasks.

% The control group participants only saw a small icon that did not disrupt their browsing experience, and their login times were therefore similar to typical times without the tool.
% Therefore, as long as website locations remain the same, \textsf{LocationWatch} did not impact user experience since the full features used in the experiment group did not lead to a significant increase in login times.
\textsf{LocationWatch} did not interrupt users in non-critical cases since both groups took similar times in the login process in Stages 1 and 2. There was a significant difference in Stage 3, during which warnings were shown to the experiment group participants, who took extra time to look up reference information or decide whether to proceed. Combined with the task completion, this showed that the tool managed to attract users' attention with the warning message.

We also found preliminary evidence that location information may have increased significance in tasks involving sensitive data.
Significantly fewer participants completed the Dropbox tasks, citing that they did not feel comfortable uploading personal information in such situations.
However, many participants still logged in despite warnings in the banking task, which our interview results showed to be the most sensitive.
It is difficult to know exactly why this occurred, but it suggests that participants interpret location information differently in different contexts.

Our evaluation also showed that \textsf{LocationWatch} was usable.
The SUS scores were good and similar for both groups, implying that the version with the location change warning was as usable as existing solutions, which primarily show the country flag (as in the control group).
%The warning was also a desired feature suggested by many participants in the control group (8 out of 22) after the experiment. 

% As \textsf{LocationWatch} has a similar purpose to certificate validation, it may be useful to incorporate it with existing certificate indicators.

%\blue{
%In this paper, we have taken a similar approach to that by Ruoti et al.~\cite{ruoti2017weighing} by conducting semi-structured interviews to identify users' security concerns and inclinations.
%Their findings on the misconceptions of TLS indicators also support the need for developing alternatives of informing users about online security.
%Through the user experiments, we identified website location as a possible factor for users in assessing their security and privacy.
%}

% \begin{table}[t]
%   \centering
%   \begin{tabular}{lcrrr}
%   \toprule
%   Condition & Stage & Mean & Median & SD \\
%   \midrule
%   \multirow{3}{*}{Control}    & 1 & 2.95 & 3 & 0.21 \\
%                               & 2 & 2.95 & 3 & 0.70 \\
%                               & 3 & 2.00 & 3 & 1.15 \\
%   \midrule
%   \multirow{3}{*}{Experiment} & 1 & 2.64 & 3 & 0.49 \\
%                               & 2 & 2.45 & 3 & 0.96 \\
%                               & 3 & 1.14 & 1 & 0.91 \\
%   \bottomrule
%   \end{tabular}
%   \caption{Descriptive statistics of task completion across different websites.}
%   \label{tab:complete_descriptive_website}
% \end{table}

\section{Discussion}
\label{sec:discussion}

The results of our studies suggest that website location is a promising research direction for helping users authenticate websites.
We discuss various aspects of using website location as a factor for users to relate to their security and privacy.

\subsection{Adoption and Deployment}
With the trend of data localization~\cite{bowman2017data}, location verification could become an important consideration for data center deployment, which is presently concerned with infrastructure and sustainability~\cite{datacenterindex}.
Companies often host their websites using content delivery networks (CDNs), which serve data from servers closer to the clients (often to the same city or region).
We envision that market and regulatory pressure would encourage companies to choose CDNs in preferred locations.
We see the beginnings of this with EV certificates~\cite{evcerts}, which contain verified information about company office locations.
The security benefits of automatic and verifiable location information would further encourage wider adoption.

For real-world deployment, website owners could opt in to provide detailed and up-to-date location information upon client connection, as a service to provide extra security assurance.
We envision that \textsf{LocationWatch} would store legitimate server locations for each supported website, which might consist of multiple locations. If the tool detects that the website is being served from an unlisted location, it would display a warning similar to that in \textsf{LocationWatch}. We found that users' security awareness could be raised by such warnings, allowing them to take further caution with their private data.

\subsection{Challenges of Location Authentication}

%We implemented \textsf{LocationWatch} as a separate tool because we wanted our participants to be aware of its existence while minimizing the influence of their preconceptions of TLS warnings. However, we envision that \textsf{LocationWatch}'s features could be integrated with existing security indicators, allowing the user to inspect website location with other security information in a consolidated manner.

One challenge with any added security mechanism is that it may distract or overwhelm users, who in response bypass or ignore warnings, and this has been addressed with ongoing research~\cite{felt2016rethinking}.
%There is ongoing research to balance catching users' attention with relevant warnings about their privacy or security without overwhelming them.
We implemented \textsf{LocationWatch} as a separate tool to minimize the influence of other TLS warnings in our studies.
However, we envision that it could be integrated with existing indicators to provide users with security information in a consolidated manner.

There are also challenges inherited from location-based authentication techniques due to the Internet infrastructure.
First, it is likely that many websites are physically hosted on the same CDN server. This allows an attacker that compromises a website to host a phishing website on the same CDN, effectively serving arbitrary content from the same location. This attack is not detectable in the currently proposed location-verification protocols~\cite{abdou2016server, yu2016salve} since the location ceases to be a unique factor of the website for verification. However, critical websites can resist such attacks if they host their login web pages on privately owned data centers. The warnings of our location indicator are effective only if the TLS endpoint of the website is served from unique and private locations.

Another limitation of location-based authentication stems from the use of third-party resources (like CSS and JavaScript), which can be served from other locations. Like current security indicators, \textsf{LocationWatch} shows the location of the top-level web server, while the locations of embedded content are not displayed. This is a design choice made to avoid confusing users, as they typically view complete webpages and do not consider individual webpage elements. Creating a design that allows users to explore the locations of individual elements in a single page is an open challenge.
% A more detailed interface which potentially provides a mapping from individual resources to their respective originating locations may be too complicated for typical web users and is left for future investigation.

Adversarial co-location and third-party resources are fundamental limitations of recent location-based authentication mechanisms.
While our exploration of location as a security factor inherits such limitations, we found that there is potential for location information as a comprehensible security indicator. 

% Users also rely on usable security tools to provide comprehensible security tools to aid their decisions.

\subsection{Study Limitations}

This type of work is unavoidably affected by participants' views.
We conducted our studies in person to obtain richer data about users' interactions with \textsf{LocationWatch} and how they perceived the warnings.
However, it limited the diversity of perspectives that we captured (both in terms of participants and the number of websites and locations we could present).
In future work, it would be interesting to conduct a larger study using crowdsourcing platforms to evaluate location's influence and a more global perspective.

Our study was also affected by the aforementioned challenges of location-based authentication.
However, the in-person experiments allowed us to interactively observe users and their diverse contingent actions (e.g., retroactively deleting files or reasoning about warnings).
%Overall, our study explored the impact of users' knowledge of website locations on their security awareness.

\section{Conclusion}
\label{sec:conclusion}

Authenticating websites is an important problem that affects users because they must make decisions about whether a website is trustworthy. The current certificate model forces users to interpret dense and unfamiliar technical information, which results in users expressing confusion about warnings or ignoring them in favor of non-technical cues~\cite{felt2015improving}. Recent proposals suggest the addition of web server location authentication~\cite{abdou2016server, yu2016salve} to strengthen TLS. Our work is the first to explore the usability aspects of these proposals. We investigated how location information can fit into users' decision-making processes, and whether location information affects the decisions that users make about security-sensitive tasks. 

We designed \textsf{LocationWatch}, a browser extension to notify users of website locations and their changes. We conducted a user study and found that when alerted to a location change, users understood the change and interpreted it in light of their current task. As a result, they were less likely to complete security-sensitive tasks when warned about location changes. Our findings suggest that website location indication has the potential to be a usable approach to helping users make informed decisions about privacy and security. 

\section{Acknowledgment}
We are grateful for Marco Guarnieri for his feedback on improving the paper.
We also thank the anonymous reviewers and shepherd for their helpful comments. 
Der-Yeuan Yu was funded by ABB Corporate Research, Switzerland.
This paper represents authors' views. 



\bibliographystyle{IEEEtranS}
\bibliography{IEEEabrv,location}
%\begin{thebibliography}{1}
%\end{thebibliography}

\appendix

\section{User Interview Script}
\label{sec:interview_script}

We attach our semi-structured interview script here.
We prepared the following interview questions to ask participants about their Internet use, location and security awareness, and preferences on location.

\textbf{Online Storage}\\
1. Do you use online file storage services (e.g., Dropbox, Apple iCloud, Google Drive)? Could you mention some examples of how you use it?

2. What kinds of data do you store online? Are there types of data that you typically try not to put on the Internet?

3. Is there any information about yourself that you specifically try not to store on the Internet? 

\textbf{Email, Calendars, Contacts}\\
4. Do you use online calendars like the Google Calendar, or the iCloud calendar? Can you elaborate on the types of events you mark on your calendar that you store online?

5. Do you use a web-based email service? What do you use it for?

6. Do you store your contact information online? What kinds of information do you store?

\textbf{Finance and Shopping}\\
7. Do you use online banking? Could you mention some examples of how you use it? E.g., just checking your balance, transferring funds, stocks investment or financial planning.

8. Do you use your credit card to shop online? 

9. How do you choose where to shop online? What kind of considerations are likely to make you trust an online store? Will it make a difference to you if you know you can access a brick-and-mortar branch of that store?

10. When you are shopping online, how would you feel if store's domain indicates a foreign country?

11. What kind of precautions do you take around handling financial transactions online (whether with credit cards or online banking)? Do you store your credit card information online?

12. Are there any aspects of financial management that you would not feel comfortable performing online?

\textbf{Social Networking}\\
13. Do you use social networking or messenger services, such as Facebook, Google Plus, Twitter, Instagram, etc.? This may also include messaging services like WhatsApp.

14. Is there a difference in the kind of information you share to different platforms? What kind of considerations do you make before putting your information on different types of social media?

15. What are your concerns regarding your privacy on social networking websites, such as Facebook, Twitter, or Instagram?

\textbf{Knowledge of Locations}\\
16. When you store your files (photos, videos, documents) online, where do you think these files are stored?

17. When you visit a website, such as Wikipedia, Google Maps, or Yahoo News, where do you think the web content is stored?

18. When you visit a website, such as online banking or online storage, how do you know you are actually visiting the real website, as opposed to a forged website to steal your personal information? Are there particular indicators that you pay attention to?

\textbf{Internet Service Location Preferences}\\
\textit{Interviewer: We've so far talked about a lot of things you can do using the Internet. A lot of these services store your data in data centers located somewhere in the world. Companies also use these data centers to store information that you consume, such as news articles. Let's talk about your trust or various preferences regarding these data centers.}

1. What are your privacy concerns about your data online? This might include files stored online, personal information, or credit cards? 

2. Do you have any concerns about where your data are being stored? What kind of concerns? For example, where would you like your data to be stored?

3. Does your preference of where your data is stored depend on the type of data? Specifically, consider the following types of data: your banking account data, online shopping history, chats, emails, social networking data, hotel or flight bookings, etc.

4. Imagine that you are provided with information regarding the location of where your online services are. How would such information influence your trust in these services?

5. What kind of location information do you have in mind? How detailed would you prefer such information to be presented? 

6. Imagine that the location information can be presented to you when you visit a website. How do you think this location information should be displayed?

7. If you had location information available to you, in what kind of services do you think it would be useful?

\section{Test Statistics}

\begin{table}[h]
	\small
	\centering
	\begin{tabular}{ccccccc}
		\toprule
		& \multicolumn{3}{c}{Control} & \multicolumn{3}{c}{Experiment} \\
		Stage & Mean & Mdn & SD & Mean & Mdn & SD \\
		\midrule
		1 & 2.82 & 3 & 0.50 & 2.73 & 3 & 0.63 \\
		2 & 2.55 & 3 & 0.60 & 2.32 & 2 & 0.65 \\
		3 & 2.32 & 3 & 0.84 & 1.45 & 2 & 0.96 \\
		\bottomrule
	\end{tabular}
	\caption{Descriptive statistics of task completion across different stages.}
	\label{tab:complete_descriptive_stage}
\end{table}

\begin{table}[h]
	\small
	\centering
	\begin{tabular}{ccccccc}
		\toprule
		& \multicolumn{3}{c}{Control} & \multicolumn{3}{c}{Experiment} \\
		Stages    &   $\chi^2$   & df &   $p$   &   $\chi^2$   & df &   $p$      \\
		\midrule
		All       &     7.35     &  2 &  0.228  &    30.86     &  2 & $<0.001$ \\ % \input{chisq_session_ctrl_all}
		S1 vs. S2 &     --      & -- &    --   &     3.62     &  1 &  0.513     \\ %\input{chisq_session_ctrl_12} 
		S1 vs. S3 &    --        & -- &   --   &    26.15     &  1 & $<0.001$   \\ % \input{chisq_session_ctrl_13}
		S2 vs. S3 &    --        & -- &   --    &    10.52     &  1 &  0.011     \\ % \input{chisq_session_ctrl_23}
		%		S1 vs. S2 &     2.00     &  1 &  1.000  &     3.62     &  1 &  0.513     \\ %\input{chisq_session_ctrl_12} 
		%		S1 vs. S3 &     6.15     &  1 &  0.118  &    26.15     &  1 & $<0.001$   \\ % \input{chisq_session_ctrl_13}
		%		S2 vs. S3 &     0.79     &  1 &  1.000  &    10.52     &  1 &  0.011     \\ % \input{chisq_session_ctrl_23}
		%   All stages & 7.35 & 2 & 0.025 & 30.86 & 2 & $<0.001$ \\ % \input{chisq_session_ctrl_all}
		\bottomrule
	\end{tabular}
	\caption{Chi-squared tests of task completion across different stages using the Bonferroni correction. We did not perform pairwise tests on the control condition since we found no significant differences between all the stages.}
	\label{tab:complete_chisq_stage}
\end{table}

\begin{table}[t]
	\small
	\centering
	\begin{tabular}{lcrcrrr}
		\toprule
		Cond                   & Stage &  Mean  & Mdn & SD    & Skew  & Kurtosis \\
		\midrule
		\multirow{3}{*}{Ctrl}  &     1 &  33.77 &  28 & 13.13 &  0.61 & -0.35    \\
		&     2 &  16.83 &  17 &  4.22 & -0.59 &  1.44    \\
		&     3 &  18.29 &  14 & 16.91 &  3.01 & 10.93    \\
		\midrule
		\multirow{3}{*}{Expt}  &     1 &  34.32 &  33 &  8.61 &  0.85 &  0.86    \\
		&     2 &  16.64 &  15 &  6.63 &  0.30 & -0.88    \\
		&     3 &  54.02 &  41 & 45.12 &  1.11 &  0.77    \\
		\bottomrule
	\end{tabular}
	\caption{Descriptive statistics of decision-making times.}
	\label{tab:login_times_descriptive}
\end{table}

\begin{table}[h]
	\small
	\centering
	\begin{tabular}{ccccccc}
		\toprule
		& \multicolumn{3}{c}{Control}    & \multicolumn{3}{c}{Experiment} \\
		Stages    &  $t$  & df &     $p$  &  $t$  & df &     $p$           \\
		\midrule
		S1 vs. S2 &  6.01 & 21 & $<$0.001 &  7.24 & 21 & $<$0.001          \\ % ttest_c_S1S2 and ttest_t_S1S2
		S1 vs. S3 &  3.52 & 21 &    0.019 & -1.74 & 21 &    0.868          \\ % ttest_c_S1S3 and ttest_t_S1S3
		S2 vs. S3 & -0.43 & 21 &    1.000 & -3.43 & 21 &    0.023          \\ % ttest_c_S2S3 and ttest_t_S2S3
		\bottomrule
	\end{tabular}
	\caption{$t$-tests of decision-making times across different stages using the Bonferroni correction.}
	\label{tab:login_times_ttest}
\end{table}

\begin{table}[t]
	\small
	\centering
	\begin{tabular}{ccccccc}
		\toprule
		& \multicolumn{3}{c}{Control} & \multicolumn{3}{c}{Experiment} \\
		Website & Mean & Mdn & SD & Mean & Mdn & SD \\
		\midrule
		Bank & 2.95 & 3 & 0.21 & 2.64 & 3 & 0.49 \\
		Facebook & 2.73 & 3 & 0.70 & 2.45 & 3 & 0.96 \\
		Dropbox & 2.00 & 3 & 1.15 & 1.14 & 1 & 0.91 \\
		\bottomrule
	\end{tabular}
	\caption{Descriptive statistics of task completion across different websites.}
	\label{tab:complete_descriptive_website}
\end{table}

\begin{table}[h]
	\small
	\centering
	\begin{tabular}{ccccccc}
		\toprule
		& \multicolumn{3}{c}{Control} & \multicolumn{3}{c}{Experiment} \\
		Tasks   &   $\chi^2$  & df &    $p$   &   $\chi^2$  & df &     $p$    \\
		\midrule
		All        &  29.17 & 2 & $<$0.001 &  32.07 & 2 & $<$0.001 \\ 
		B vs. FB   &   2.41 &  1 &    0.962 &   0.53 &  1 &    1.000   \\  % chisq_website_ctrl_bank_fb.tex chisq_website_expt_bank_fb.tex
		B vs. DB   &  21.06 &  1 & $<$0.001 &  23.32 &  1 & $<$0.001   \\  % chisq_website_ctrl_bank_db.tex chisq_website_expt_bank_db.tex
		FB vs. DB   &  10.20 &  1 &    0.011 &  15.99 &  1 & $<$0.001   \\  % chisq_website_ctrl_fb_db.tex chisq_website_expt_bank_fb_db.tex
		%   All        &  29.17 & 2 & $<$0.001 &  32.07 & 2 & $<$0.001 \\ 
		\bottomrule
	\end{tabular}
	\caption{Chi-squared tests of task completion across different websites using the Bonferroni correction.}
	\label{tab:complete_chisq_website}
\end{table}

\end{document}